\documentclass[11pt]{article}
\usepackage{color}
\usepackage{amsfonts,amssymb, graphicx,bm}
\usepackage{a4}
\usepackage[english]{babel}

\newtheorem{theorem}{Theorem}[section]
\newtheorem{proposition}[theorem]{Proposition}

\begin{document}

\def\reff#1{(\protect\ref{#1})}

\let\a=\alpha \let\b=\beta \let\ch=\chi \let\d=\delta \let\e=\varepsilon
\let\f=\varphi \let\g=\gamma \let\h=\eta    \let\k=\kappa \let\l=\lambda
\let\m=\mu \let\n=\nu \let\o=\omega    \let\p=\pi \let\ph=\varphi
\let\r=\rho \let\s=\sigma \let\t=\tau \let\th=\vartheta
\let\y=\upsilon \let\x=\xi \let\z=\zeta
\let\D=\Delta \let\F=\Phi \let\G=\Gamma \let\L=\Lambda \let\Th=\Theta
\let\O=\Omega \let\P=\Pi \let\Ps=\Psi \let\Si=\Sigma \let\X=\Xi
\let\Y=\Upsilon

\global\newcount\numsec\global\newcount\numfor
\gdef\profonditastruttura{\dp\strutbox}
\def\senondefinito#1{\expandafter\ifx\csname#1\endcsname\relax}
\def\SIA #1,#2,#3 {\senondefinito{#1#2}
\expandafter\xdef\csname #1#2\endcsname{#3} \else
\write16{???? il simbolo #2 e' gia' stato definito !!!!} \fi}
\def\etichetta(#1){(\veroparagrafo.\veraformula)
\SIA e,#1,(\veroparagrafo.\veraformula)
 \global\advance\numfor by 1
 \write16{ EQ \equ(#1) ha simbolo #1 }}
\def\etichettaa(#1){(A\veroparagrafo.\veraformula)
 \SIA e,#1,(A\veroparagrafo.\veraformula)
 \global\advance\numfor by 1\write16{ EQ \equ(#1) ha simbolo #1 }}
\def\BOZZA{\def\alato(##1){
 {\vtop to \profonditastruttura{\baselineskip
 \profonditastruttura\vss
 \rlap{\kern-\hsize\kern-1.2truecm{$\scriptstyle##1$}}}}}}
\def\alato(#1){}
\def\veroparagrafo{\number\numsec}\def\veraformula{\number\numfor}
\def\Eq(#1){\eqno{\etichetta(#1)\alato(#1)}}
\def\eq(#1){\etichetta(#1)\alato(#1)}
\def\Eqa(#1){\eqno{\etichettaa(#1)\alato(#1)}}
\def\eqa(#1){\etichettaa(#1)\alato(#1)}
\def\equ(#1){\senondefinito{e#1}$\clubsuit$#1\else\csname e#1\endcsname\fi}
\let\EQ=\Eq
\def\0{\emptyset}

\def\pp{{\bm p}}\def\pt{{\tilde{\bm p}}}


\def\\{\noindent}
\let\io=\infty

\def\VU{{\mathbb{V}}}
\def\EE{{\mathbb{E}}}
\def\GI{{\mathbb{G}}}
\def\TT{{\mathbb{T}}}
\def\C{\mathbb{C}}
\def\CC{{\mathcal C}}
\def\II{{\mathcal I}}
\def\LL{{\cal L}}
\def\RR{{\cal R}}
\def\SS{{\cal S}}
\def\NN{{\cal N}}
\def\HH{{\cal H}}
\def\GG{{\cal G}}
\def\PP{{\cal P}}
\def\AA{{\cal A}}
\def\BB{{\cal B}}
\def\FF{{\cal F}}
\def\v{\vskip.1cm}
\def\vv{\vskip.2cm}
\def\gt{{\tilde\g}}
\def\E{{\mathcal E} }
\def\I{{\rm I}}
\def\rfp{R^{*}}
\def\rd{R^{^{_{\rm D}}}}
\def\ffp{\varphi^{*}}
\def\ffpt{\widetilde\varphi^{*}}
\def\fd{\varphi^{^{_{\rm D}}}}
\def\fdt{\widetilde\varphi^{^{_{\rm D}}}}

\def\pfp{\Pi^{*}}
\def\pd{\Pi^{^{_{\rm D}}}}
\def\pbfp{\Pi^{*}}
\def\fbfp{{\bm\varphi}^{*}}
\def\fbd{{\bm\varphi}^{^{_{\rm D}}}}
\def\rfpt{{\widetilde R}^{*}}

\def\tende#1{\vtop{\ialign{##\crcr\rightarrowfill\crcr
              \noalign{\kern-1pt\nointerlineskip}
              \hskip3.pt${\scriptstyle #1}$\hskip3.pt\crcr}}}
\def\otto{{\kern-1.truept\leftarrow\kern-5.truept\to\kern-1.truept}}
\def\arm{{}}
\font\bigfnt=cmbx10 scaled\magstep1

\newcommand{\card}[1]{\left|#1\right|}
\newcommand{\und}[1]{\underline{#1}}
\def\1{\rlap{\mbox{\small\rm 1}}\kern.15em 1}
\def\ind#1{\1_{\{#1\}}}
\def\bydef{:=}
\def\defby{=:}
\def\buildd#1#2{\mathrel{\mathop{\kern 0pt#1}\limits_{#2}}}
\def\card#1{\left|#1\right|}
\def\proof{\noindent{\bf Proof. }}
\def\qed{ \square}
\def\trp{\mathbb{T}}
\def\trt{\mathcal{T}}
\def\Z{\mathbb{Z}}
\def\be{\begin{equation}}
\def\ee{\end{equation}}
\def\bea{\begin{eqnarray}}
\def\eea{\end{eqnarray}}

\def\v{\vskip.1cm}
\def\vv{\vskip.2cm}
\def\gt{{\tilde\g}}
\def\E{{\mathcal E} }
\def\I{{\rm I}}
\def\GI{\mathbb{G}}
\def\EE{\mathbb{E}}\def\VU{\mathbb{V}}

\newcommand{\nbhd}[1]{\Gamma\left(#1\right)}
\newcommand{\nbhdst}[1]{\Gamma^*\left(#1\right)}
\newcommand{\eee}{{\rm e}}
\newcommand{\colo}{\mathcal C}
\definecolor{Red}{cmyk}{0,1,1,0}
\def\red{\color{Red}}

\title {On the convergence of cluster expansions for polymer gases}

\author{Rodrigo Bissacot$^{1,2}$, Roberto Fern\'{a}ndez$^{2,3}$,  Aldo Procacci$^1$ \\\\
\footnotesize{$^1$Dep. Matem\'atica-ICEx, UFMG, CP 702
Belo Horizonte - MG, 30161-970 Brazil}
\\
\footnotesize{$^2$Labo. de Maths Raphael Salem, Universit\'e de Rouen, 76801,  France}\\
\footnotesize{$^3$Department of Mathematics, Utrecht University, P.O. Box 80010 3508 TA Utrecht}\\
\tiny{emails: {rodrigo.bissacot@gmail.com};~ {
Roberto.Fernandez@univ-rouen.fr};~ {R.Fernandez1@uu.nl};}\\
\tiny{
{ aldo@mat.ufmg.br}}
}
\date{}
\maketitle
\begin{abstract}
We compare the different convergence criteria available for cluster expansions of polymer gases subjected to hard-core exclusions, 
with emphasis on polymers defined as finite subsets of a countable set (e.g.\  contour expansions and more generally high- 
and low-temperature expansions).
In order of increasing strength, these criteria are:  (i) Dobrushin criterion, obtained by a simple inductive argument; (ii) 
Gruber-Kunz criterion obtained through the use of Kirkwood-Salzburg equations, and (iii) a criterion obtained by two of us 
via a direct combinatorial handling of the terms of the expansion.  We show that for subset polymers our sharper criterion 
can be proven both by a suitable adaptation of Dobrushin inductive argument and by an alternative ---in fact, more elementary--- 
handling of the Kirkwood-Salzburg equations.  In addition we show that for general abstract polymers this alternative treatment 
leads to the same convergence region as the inductive Dobrushin argument and, furthermore, 
to a systematic way to improve bounds on correlations.
\end{abstract}

\BOZZA
\numsec=1\numfor=1
\section{Introduction}

The most basic and frequent  applications of
cluster expansions deal with (log of) partition functions of random geometric objects subjected only to hard-core exclusions.  The relevant mathematical structure was first formalized by Gruber and Kunz~\cite{GK} for the case of objects defined by subsets of a countable set.  They called these objects \emph{polymers}.  A decade later Koteck\'y and Preiss~\cite{KP} introduced more general families of objects that are not necessarily subsets of an underlying set and whose ``hard-core'' interaction is defined by an \emph{incompatibility relation}.  This more general objects will be called \emph{abstract polymers}  in the sequel, while
 \emph{subset polymers} will be those introduced by Gruber and Kunz.

In their seminal paper, Gruber and Kunz used Kirkwood-Salzburg equations to determine convergence radii.  This GK approach, however, did not become popular and instead practitioners turned to bounds obtained by first showing that cancellations yield a majorizing expansion in terms of tree diagrams.  The convergence condition is proved by inductively  summing the ``leaves'' of the expansion.    The genesis of this approach is attributed to Cammarota~\cite{Ca}, and the canonical reference is the excellent review by Brydges~\cite{Br} (see also \cite{BM}, \cite{PdLS}, \cite{PS}).  In contrast, Koteck\'y and Preiss introduced an inductive argument that does not make any reference to the actual expression of the series.  This argument, helped by a refinment by Dobrushin~\cite{Dob}, became the argument-of-choice in further developments~\cite{narolizah99,BZ,Mi,So,uel04,SS}.  There are good reasons for this: the inductive argument leads to notoriously simpler convergence proofs and stronger results than the more laborious tree sums (see~\cite{Dob2} for a remarkable overview of consequences of these results).  In particular it leads quite naturally to a bound on correlationns and ``pinned'' free energies.  But there is a downside: the argument works ``too well".  While Dobrushin's condition is perfectly designed to survive the inductive step, the method contains no hint on how to obtain further improvements.

To break this impasse, in~\cite{FP} we went back to basics.  We took a hard look at the cluster expansion in full and studied it avoiding inequalities as much as possible.  The breakthrough came from a seldomly remembered paper by Oliver Penrose~\cite{Pe} where the series is written in terms of a tree-grap \emph{identity} involving trees determined by compatibility constraints.  This yields a series whose partial sums can be generated as successive applications of a fixed transformation.  Convergence criteria are then found by suitably bounding this transformation.  In this way we were able to improve Dobrushin criterion and, furthermore, explain the loss of precision of preexisting criteria
as incomplete accounting of Penrose constraints. Thus, on the positive side we managed to clarify the role of different approximations and to obtain stronger results, while leaving in the background a tree expansion that could be used for further refinments (see e.g. \cite{pr,pr2,pu}).  On the negative side, however, our method lacks the elegance and economy of the inductive Koteck\'y-Preiss-Dobrushin (KPD) approach.  We have already exploited the positive side
to improve results in a number of well-studied applications~\cite{bfps,fp2,fps,jps}.  The present paper addresses the negative side.

In~\cite{FP} we also  applied our new  criterion to subset polymers, resorting to some simplifications to obtain a expression suitable for computations.  We were surprised to find out that the ``new'' resulting criterion while being clearly better than Dobrushin's is in fact identical to the long forgotten Gruber-Kunz bound.  There is only a small difference in our favor: the GK bound involve a strict inequality while in ours the inequality is not strict.  That is why we call our bound the \emph{extended GK criterion}. This state of affairs motivated a number of questions that sufaced repeatedly in discussions with our colleagues and motivated the PhD thesis of one of us~\cite{Bis}:

\begin{itemize}

\item[(Q1)] Does the use of Kirkwood-Salzburg equations \emph{\`a la} Gruber-Kunz yield better bounds than the inductive KPD approach?

\item[(Q2)] Can Kirkwood-Salzburg equations lead to an alternative proof of our new criterion?

\item[(Q3)]  More ambitiously:  Is there an inductive proof ---\emph{\`a la} Dobrushin--- of our new criterion?

\end{itemize}

In this paper we answer these questions for the case of subset polymers, and provide partial answers for general abstract polymers.  In more detail we prove the following:

\begin{itemize}

\item[(A1)] For subset polymers:

\begin{itemize}

\item[(A1.1)] An alternative handling of the Kirkwood-Salzburg equations proves the extended GK criterion.  Thus,  Kirkwood-Salzburg equations can indeed be used to prove our new convergence criterion and bounds on correlations for these polymers.  The alternative handling consists in replacing the Banach-space fixed-point theorem of the original approach by a more elementary argument on convergence of series with positive terms.  The extended criterion follows from writing this expansion as a limit of iterations of a fixed transformation, following an idea from~\cite{FP}.


\item[(A1.2)] Suitably adapted, an inductive KPD-approach can also be used to prove the extended GK criterion.  Therefore, for subset polymers the three approaches ---GK, inductive KPD and ours--- are equivalent.

\end{itemize}

\item[(A2)] Likewise, the GK approach with modified handling of the Kirkwood-Salzburg equations is equivalent to the inductive KPD approach for general abstract polymers.

\end{itemize}

In the general setup of abstract polymers, we are at present unable to prove our improved criterion using either of these two equivalent approaches. The obstacle, explicitly seen in our treatment below, is the use of \emph{factorized} majorizing weights inherent to both the GK and the KPD approaches.

The paper is organized so to be reasonably self-contained.   In Section 2 we review the general definition of abstract and subset polymer gases and we present and compare the different convergence criteria.  In Section 3 we show how to obtain the (extended) Gruber-Kunz criterion through an inductive argument.  We adapt a simple argument by Miracle-Sol\'e~\cite{Mi} that relies on the alternating-sign property of the truncated coefficients. In Section 4  we review Gruber-Kunz setting of Kirkwood-Salzburg equations for subset polymers and prove the extended GK bound by introducing the alternative treatment of these equations.   In Section 5 we show how Dobrushin criterion can also be obtained from Kirkwood-Salzburg equations in the general abstract case.  We conclude with some final comments and suggestions.

\section{Polymer gases and convergence criteria}
\numsec=2\numfor=1

First, some general notation.  For a set $U$ we denote $|U|$ its cardinality and $\ind{U}$ its indicator function.

\subsection{The abstract polymer gas}

An abstract polymer system is a triple $(\PP,\mathcal{R}, \bm z)$ where
\begin{itemize}

\item $\PP$ is a countable set, whose  elements $\g$  we call polymers, following Gruber and Kunz.

\item $ \mathcal{R}\subset\PP\times \PP$ is a  {\it symmetric and reflexive relation}.
When $(\g,\g')\in \mathcal{R}$, we write $\g\not\sim \g'$ and say that $\g$ and $\g'$ are {\it incompatible}. Conversely, if
$(\g,\g')\notin \mathcal{R}$ we say that the polymers $\g$ and $\g'$ are
{\it compatible} and we write $\g\sim\g'$. Note that  the assumption that $ \mathcal{R}$ is reflexive
implies that $\g\not\sim\g$ for all $\g\in \PP$.

\item $\bm z:\PP\to \C: \g \mapsto z_\g$ is the \emph{activity function}.
The number $z_\g$ is  called the activity of the  polymer $\g$.

\end{itemize}

The corresponding polymer gas is defined by complex-valued measures.
For each finite family of polymers $\L\subset \PP$ a measure is defined assigning, to each polymer configuration $\{\g_1,\dots, \g_n\}\subset \L$, $n\ge 0$, the weight
$$
\mathbb P_\L(\bm z, \g_1,\dots, \g_n)={1\over \Xi_{\L}(\bm z)}~ z_{\g_1}z_{\g_2}\dots z_{\g_n}
\prod_ {1\leq i<j\leq n}
\ind{\g_i\sim\g_j}\Eq(proba)
$$
with the convention $\{\g_1,\dots, \g_n\}=\emptyset$ when $n=0$ and $\mathbb P_\L(\bm z, \emptyset)={1/\Xi_{\L}(\bm z)}$.  Here
$$
\begin{minipage}[c]{9cm}
\begin{eqnarray*}
\Xi_{\L}(\bm z)&=& 1+\sum_{n\geq 1}{1\over n!}
\sum_{(\g_{1},\dots ,\g_{n})\in\L^n}
z_{\g_1}z_{\g_2}\dots z_{\g_n}
\prod_ {1\leq i<j\leq n}
\ind{\g_i\sim\g_j}\\
&=& 1+\sum_{n\geq 1}
\sum_{\{\g_{1},\dots ,\g_{n}\}\in\L}
z_{\g_1}z_{\g_2}\dots z_{\g_n}
\prod_ {1\leq i<j\leq n}
\ind{\g_i\sim\g_j}
\end{eqnarray*}
\end{minipage}
\Eq(2)
$$
normalizes $\mathbb P_\L(\bm z, \L)=1$.  The equality betwwen the first and second lines follows from the fact that, in the presence of compatibility requirements, only $n$-t-uples with different components contribute.
Restricted to positive fugacities,,
 $\{z_\g\ge 0\}_{\g\in \PP}$,  the measure $\mathbb P_\L$ is a probability measure on the space of subsets of $\L$ and $\mathbb P_\L(\bm z, \g_1,\dots, \g_n)$ is interpreted as
the probability of observing exactly polymers $\g_1,\dots, \g_n$ out of the family $\Lambda$.   In general, weights are allowed to be complex to settle analyticity questions.
The normalization constant $\Xi_{\L}(\bm z)$ is interpreted as
the grand-canonical partition function of the family $\L$.
This partition function $\Xi_{\L}(\bm z)$  is the key function from which all ``physical quantities'' of the system can be derived. These quantities include
the  ``pressure'' of the system
$$
P_\L(\bm z)={1\over |\L|}\log \Xi_{\L}(\bm z)\Eq(pressure)
$$
and the correlations
$$
\phi_\L(\bm z,\g_1,\dots, \g_p)\;=\; \Bigl[z_{\g_1}\dots z_{\g_p}\,\prod_{1\le i<j\le p}\ind{\g_i\sim\g_j}\Bigr]
 {\Xi_{\L\backslash \nbhd{\cup_{i=1}^p \g_i}}(\bm z)\over\Xi_\L(\bm z)}\;.
 \Eq(correl)
$$
assuming $\{\g_1,\dots, \g_p\}\subset\L$ and denoting, for any finite family of polymers $X$,
$$
\nbhd{X} \;=\; \bigl\{\g\in\PP:\exists \g'\in X \mbox{ such that }\g\not\sim\g'\bigr\}
 \Eq(nbhd)
$$
---the \emph{neighborhood} of $X$.  We see from \equ(correl) that all is decided  by the ratios
$$
\bar \phi_\L(\bm z, X)\;=\;
{\Xi_{\L\backslash X}(\bm z)\over\Xi_\L(\bm z)}
\qquad\qquad X\subset\Lambda\Eq(sveg2-r)
$$
which { Gruber and Kunz baptized} the \emph{reduced correlations}. 

The main issues of the theory are the existence and analyticity with respect to fugacities of the $\L\to\PP$ limits of the pressure and correlation functions.  The most detailed way of answering these questions is by writing $\log \Xi_{\L}(\bm z)$  as a formal power series in the fugacities.  This series, historically called Mayer series, takes the form
$$
\log\Xi_{\L}(\bm z)\;=\; \sum_{n=1}^{\infty}{1\over n!}
\sum_{(\g_{1},\dots ,\g_{n})\in\L^n}
\phi^{T}(\g_1 ,\dots , \g_n)\,z_{\g_1}z_{\g_2}\dots z_{\g_n}\Eq(6)
$$
where the \emph{truncated coefficients} $\phi^{T}(\g_{1},\dots ,\g_{n})$ depend only on the compatibility graph of the argument.  This graph has vertex set $\{1,2,\dots,n\}$ and edge set
$\bigl\{\{i,j\}\subset \{1,2,\dots,n\}: \g_i\nsim\g_j \bigr\}$.   [\emph{Warning!} The truncated coefficients do not require compatibility of their arguments, hence the analogous of the second line of \equ(2) is \emph{not} valid for \equ(6).]
The methods discussed in this paper do not make use of the actual expression of the truncated coefficients (which can be found, for instance, in~\cite{Ca,Br, BM, PdLS,PS}).  Below we only need the \emph{alternating-sign property}
$$
\bigl|\phi^{T}(\g_{1},\dots ,\g_{n})\bigr|\;=\; (-1)^{n-1}\phi^{T}(\g_{1},\dots ,\g_{n})\;,
\Eq(sign)
$$
which can be easily derived from the Penrose identity~\cite{Pe} (see e.g.~\cite{FP} and \cite{SS}) or by a simple induction argument~\cite{Mi}.

In view of \equ(sveg2-r) it is natural to focus on the differences
$$\begin{array}{rcl}
\Theta^\L_{\g}(\bm z)& = & \log \Xi_{\L}(\bm z)- \log \Xi_{\L\backslash\{\g\}}(\bm z)\cr\cr
&=& \displaystyle{\sum_{n=1}^{\infty}{1\over n!}
\mathfrak{}\sum_{(\g_{1},\dots ,\g_{n})\in\L^n\atop \exists i:~ \g_i=\g}
\phi^{T}(\g_1 ,\dots , \g_n)\;{z_{\g_1}}\dots{z_{\g_n}}}\;.
\end{array}
\Eq(diflog-g)
$$
From them we can reconstruct
$$
\log\Xi_{\L}(\bm z)\;=\; \sum_{\g\in\L}\Theta^\L_\g(\bm z)
\Eq(reconst)
$$
and, for $\{\g_1,\ldots,\g_p\}\subset \L$,
$$
\bar \phi_{\L}\bigl(\bm z, \{\g_1,\ldots,\g_p\}\bigr)\;=\;
\exp\Bigl(-\sum_{i=1}^{p}\Theta^{\L\backslash\{\g_{i+1},\ldots,\g_p\}}_{\g_i}(\bm z)\Bigr)\;.
\Eq(recons-x)
$$

{
In order to state the theorems of  the next section, we will also need to consider another series, directly related to $\Theta^\L_{\g}(\bm z)$, namely

$$\begin{array}{rcl}
\Pi^\L_{\g}(\bm z)&=& \displaystyle{{\partial\over \partial z_{\g_0}}\log \Xi_{\L}(z_\L)}\cr\cr
&=&\displaystyle{\sum_{n=0}^{\infty}{1\over n!} \sum_{(\g_1,\g_2,\dots,\g_n)\in \L^n}
\phi^{T}(\g,\g_1 ,\dots , \g_n){z_{\g_1}}\dots{z_{\g_n}}}
\end{array}\Eq(Pi)
$$
}

Series \equ(6), \equ(diflog-g) { and \equ(Pi)} are examples of \emph{cluster expansions}; we will focus on the last two.   Analyticity results are obtained from it on the basis of the observation that,
for a family of positive numbers $\{\r_\g\}_{\g\in \PP}$ , the positive-term series
$$
\card{\Theta}^{\L}_\g(\bm\r)\;=\;
\sum_{n=1}^{\infty}{1\over n!}
\sum_{(\g_{1},\dots ,\g_{n})\in \L^n\atop \exists i:~ \g_i=\g}
\card{\phi^{T}(\g_1 ,\dots , \g_n)}\,\r_{\g_1}\cdots{\r_{\g_n}}
\Eq(6abs)
$$
$$
|\Pi|^\L_{\g}(\bm \r)
\;=\;\displaystyle{\sum_{n=0}^{\infty}{1\over n!} \sum_{(\g_1,\g_2,\dots,\g_n)\in \L^n}
|\phi^{T}(\g,\g_1 ,\dots , \g_n)|\,{\r_{\g_1}}\dots{\r_{\g_n}}}
\Eq(Pia)
$$
dominates \equ(diflog-g) term-by-term for $\card{z_\g}\le\r_\L$.  Therefore, convergence of this last series implies
the absolute and uniform convergence of \equ(diflog-g) { and \equ(Pi)} in the polydisc
$$\mathcal D_{\bm\r}=\bigl\{\bm z: \card{z_\g}\le\r_\g\bigr\}
\Eq(polyd)
$$
and its analyticity (with respect to the fugacities) in its interior. In fact, the alternating-sign property \equ(sign) implies that
$$
\card{\Theta}^{\L}_\g(\bm\r)=- \Theta^{\L}_\g(\bm z=-\bm \r)\;.\Eq(minus)
$$
$$
\card{\Pi}^{\L}_\g(\bm\r)= \Pi^{\L}_\g(\bm z=-\bm \r)\;.\Eq(minusa)
$$
Thus, for finite $\Lambda$, the convergence of the series \equ(6abs) { and \equ(Pia)}
is a necessary and sufficient condition for the convergence of the cluster expansion \equ(diflog-g) { and
\equ(Pi)} in the polydisc $\mathcal D_{\bm\r}$.  By \equ(reconst), \equ(recons-x) and \equ(Pi) these properties are inherited by  the correlations and the pressure.  To extend these existence and analyticity results to the limit $\L\to\PP$  the strategy is to prove that the convergence of \equ(6abs) happens for \emph{$\L$-independent} values of $\r_\L$.

\subsection{Convergence criteria for abstract polymer gases}
We focus on two criteria.  First, Dobrushin's:

\begin{theorem}{\bf \cite{Dob}}\label{Dob}
Let  ${\bm \mu}=\{\mu_\g\}_{\g\in {\PP}}$  and ${\bm \r}=\{\r_\g\}_{\g\in {\PP}}$ be  collections of nonnegative numbers such that
$$
\r_\g\, \varphi^{\rm D}_\g(\bm \mu)\;\le\;\mu_\g \qquad\qquad\forall \g\in {\PP}
\Eq(do)
$$
with
$$
\varphi^{\rm D}_\g(\bm \mu)\;=\; \prod\limits_{\gt\in\nbhd{\g}}(1+\m_\gt)\;, \Eq(cridob)
$$
then  the  series  $\card{\Theta}^{\L}_\g(\bm\r)$, $\card{\Pi}^{\L}_\g(\bm\r)$ defined in  \equ(6abs), \equ(Pia) are convergent and furthermore
$$
|\Pi|^\L_{\g}(\bm \r)\;\le \;\varphi^{\rm D}_\g(\bm \mu) \Eq(bdo)
$$
$$
\card{\Theta}^{\L}_\g(\bm\r)\;\le\; \log (1+ \mu_\g)\;.
\Eq(do.bo)
$$
\end{theorem}

We denote $\nbhd{\g}\equiv\nbhd{\{\g\}}$ the neighborhood of $\g$, namely
$\nbhd{\g}=\bigl\{\gt\in\PP: \gt\not\sim\g\bigr\}$.  Here is the second criterion, which  improves Dobrushin's.

\begin{theorem}{\bf \cite{FP}}\label{FPn}
Let  ${\bm \mu}=\{\mu_\g\}_{\g\in {\PP}}$  and ${\bm \r}=\{\r_\g\}_{\g\in {\PP}}$ be  collections of nonnegative numbers such that
$$
\r_\g\, \varphi^{\rm FP}_\g(\bm \mu)\;\le\;\mu_\g \qquad\qquad\forall \g\in {\PP}
\Eq(fp)
$$
with
$$
\begin{minipage}{10cm}
\begin{eqnarray*}
\varphi^{\rm FP}_\g(\bm \mu)&=& \Xi_{\nbhd\g}(\bm \mu)\\
&=&
\displaystyle{1+\sum_{n\ge 1}\frac{1}{n!} \sum_{(\g_1,\dots,\g_n)\in \PP^n}\!\!\!\!\!\m_{\g_1}\dots\m_{\g_n}\prod_ {1\leq i<j\leq n}\!\!\!\!\ind{\g_i\sim\g_i}\,\prod_ {i=1}^n\ind{\g_i\not\sim\g}}
\end{eqnarray*}
\end{minipage}
\Eq(fppn)
$$
then  the  series  $\card{\Theta}^{\L}_\g(\bm\r)$, $\card{\Pi}^{\L}_\g(\bm\r)$ defined in  \equ(6abs), \equ(Pia) are convergent and furthermore

$$
|\Pi|^\L_{\g}(\bm \r)\;\le \;\varphi^{\rm FP}_\g(\bm \mu) \Eq(bPia)
$$
$$
\card{\Theta}^{\L}_\g(\bm\r)\;\le\: -\ln(1-\r_\g)^{\varphi^{\rm FP}_\g(\bm \mu)-\m_\g}
\Eq(lovasz)
$$
\end{theorem}

The form \equ(do) of Dobrushin criterion shows more clearly the improvement brought by Theorem~\ref{FPn}.
In particular, since $\varphi^{\rm D}_\g(\bm \mu)\ge \varphi^{\rm FP}_\g(\bm \mu) $, for any fixed $\g, \bm \mu$, the convergence radius
$\bm R^{\rm FP}=\{\m_\g/\varphi^{\rm FP}_\g(\bm \mu)\}_{\g\in \PP}$ given by theorem \ref{FPn} is always greater than the
the  Dobrushin's crtiterion convergence radius $\bm R^{\rm D}=\{\m_\g/\varphi^{\rm D}_\g(\bm \mu)\}_{\g\in \PP}$
(the $\bm\mu$'s  here are free parameters
that can be adjusted to maximize the radii $\bm R^{\rm D}, \bm R^{\rm FP}$). The upper bound \equ(lovasz) is not
explicitly given in reference \cite{FP}. It can be proven, however (see \cite{bfps}), in a straightforward
way from \equ(bPia).
Moreover it is easy to see (see again \cite{bfps})  that \equ(lovasz)  is an improvment of \equ(do.bo) for any  $\bm \r\le \bm R^{\rm D}$.

The original statement by Dobrushin is obtained by substituting $\mu_\g +1= \eee^{a_\g}$.  In terms of these exponential weights the criterion is the existence of positive numbers
${\bm a}=\{a_\g\}_{\g\in {\PP}}$  and ${\bm \r}=\{\r_\g\}_{\g\in {\PP}}$ such that
$$
\r_{\g}\;\le\;  \Bigl(\eee^{\alpha_{\g}}-1\Bigr)\,
\exp\Bigl(-\sum_{\gt\in\nbhd{\g}} \alpha_\gt\Bigr)
\Eq(r.dob.1)
$$
and the bound \equ(do.bo) becomes:
$$
\card{\Theta}^{\L}_\g(\bm\r)\;\le\; a_\g\;.
\Eq(r.dob.2)
$$

The earlier Koteck\'y-Preiss criterion~\cite{KP} was the first to take the form \equ(do), but with $\varphi^{\rm D}_\g$ replaced by the less efficient
$$
\varphi^{\rm KP}_\g \;=\; \exp\Bigl\{\sum_{\gt\in\nbhd{\g}} \mu_\g\Bigr\}\;.
$$
The usual form of this condition, obtained upon substituting $\mu_\g = \r_\g\,\eee^{a_\g}$, is
$$
\sum_{\gt\in\nbhd{\g}} \r_\g\,\eee^{a_\g}\;\le\; a_\g\;.
\Eq(kp-abs)
$$

\subsection{Subset gases: definition and convergence criteria}
Subset gases are particular types of
polymer gases that appear in most of the uses of the cluster expansion in
statistical mechanics. Their definition requires a countable set $\VU$ that acts as an underlying ``space''.  Polymers are then simply defined as the finite non empty subsets of  $\VU$, i.e.
$$
\PP=\{\g\subset \VU : 0< |\g|<\infty\}
$$
with non-empty intersection as incompatibility relation:
$$
\g\not\sim\g'\,\,\,\,\Longleftrightarrow\,\,\,\, \g\cap\g'\neq\0\;.
\Eq(compa-r)
$$
Polymers can now be measured through its cardinality, so it makes sense
about large and small polymers. The definition of the gas is completed by a family of activities $\bm z=\{z_\g\in \C\}_{\g\in \PP}$.

In fact, the following discussion applies, more generally, to \emph{colored subset gases}.  These are systems in which polymers are endowed with some further attribute ---the \emph{color}--- taken from some space $\colo$.  Formally, colored polymers are pairs $\g=(\underline\g, c)$ with $\underline\g\subset\VU$ finite ---the \emph{support} of $\g$--- and $c\in\colo$.  Conspicuous examples are the ``thick'' contours of Pirogov-Sinai theory (see e.g.~\cite[Chapter II]{sin82}) in which colors correspond to configurations on $\underline\g$. Colors do not play any role in the incompatibility relation, which remains as in \equ(compa-r) but involving supports, that is
$\g\not\sim\g'\,\Longleftrightarrow\, \underline\g\cap\underline \g'\neq\0$.  The following expressions remain valid for these more general contours if each $\g$ is identified with its support in geometrical statements (e.g.\ $\g\in\Lambda$, $x\in\g$, etc.)

For subset gases, the different objects of interest refer to parts of the underlying $\VU$. Thus, subsets of $\VU$ both are polymers and determine ``finite-window'' magnitudes.  Because of their geometrical interpretation, subsets playing the latter role will be called ``regions''.  The corresponding definitions ---analogous to those for abstract polymers but with a slight and natural change in notation--- are as follows.
For a finite region $\Lambda\subset\VU$ and contours $\g_1,\ldots\g_n\subset\Lambda$, the probability weights $\mathbb P_\L$ are defined as in \equ(proba) with grand-canonical partition function
$$
\Xi_{\L}(\bm z)\;=\;1+\sum_{n\ge
1}\sum_{\{\g_1,\dots,\g_n\}\subset \L\atop |\g_i|\ge
1,~~\g_i\cap\, \g_j=\0} z_{\g_1}\dots
 z_{\g_n}\Eq(Xis)
$$
With these partition functions, the pressure is defined also by \equ(pressure) but for the correlations \equ(correl) we have the simplification that
$\ \nbhd{\cup_{i=1}^p \g_i}$ is replaced by $\cup_{i=1}^p \g_i$.  Therefore the reduced correlations take the form \equ(sveg2-r) for the partitions \equ(Xis).  The analogous of \equ(diflog-g) are
$$\begin{array}{rcl}
\Theta^\L_{x}(\bm z)& = & \log \Xi_{\L}(\bm z)- \log \Xi_{\L\backslash\{x\}}(\bm z)\cr\cr
&=& \displaystyle{\sum_{n=1}^{\infty}{1\over n!}
\sum_{(\g_{1},\dots ,\g_{n})\in\L^n\atop \exists i:~ \g_i\ni x}
\phi^{T}(\g_1 ,\dots , \g_n)\;{z_{\g_1}}\dots{z_{\g_n}}}
\end{array}
\Eq(diflog-g-sg)
$$
for $x\in\Lambda$.
The reconstruction formulas \equ(reconst)--\equ(recons-x) hold with $\g_i\to x_i$.
{
Note also that in the specific case of the subset
gas the series $\Theta^\L_{x}(\bm z)$ and $\Pi_\L^{\{x\}}(\bm z)$ are  in a very simple relation. Indeed,
$$
\Pi_\L^{\{x\}}(\bm z)= {\partial\over \partial z_{\{x\}}}\log \Xi_{\L}(\bm z) = {1\over  \Xi_{\L}(\bm z)}
{\partial\;\Xi_{\L}(\bm z)\over \partial z_{\{x\}}} = {\Xi_{\L\backslash\{x\}}(\bm z)\over  \Xi_{\L}(\bm z)}= \exp\{ -{\Theta}^{\L}_x(\bm z) \}
$$
whence, recalling  \equ(minus) and \equ(minusa), we get the identity
$$
|\Pi|_\L^{\{x\}}(\bm z)=\exp\{ \card{\Theta}^{\L}_x(\bm\r) \}\;.\Eq(rela)
$$
}
The convergence criteria for these polymers involve factorized weights of the form
$$
\mu_\g\;=\;\prod_{x\in\g} \xi_x\;\equiv\;{\bm \xi}^\g \Eq(fact)
$$
for some family $\bm \xi=\{\xi_x\}_{x\in\VU}$ with each $\xi_x>0$.  These single-site weights are always larger than one
and are traditional parametrized as $\xi_x=\eee^{a_x}$ with $a_x>0$.  In the literature, the $a_x$ are invariably chosen independent of $x$ and equal to some common value $a$.  With this choice
$$
\mu_\g \;=\; \eee^{a\card\g}\;.
\Eq(choice)
$$
In the sequel, however, we work with
the more general choice \equ(fact).  This introduces a minimal notational cost, but it reveals more clearly the essence of the different arguments and leads to more precise formulas.

The most widely used criterion is, in fact, Koteck\'y and Preiss' \equ(kp-abs), written in the form
$$
\sup_{x\in \VU}\,\sum_{\g\in \PP\atop\g\ni x} \,\r_\g\, \eee^{a|\g|}\;\le \; a\;.\Eq(kps)
$$
Dobrushin criterion (Theorem~\ref{Dob}) with the substitution
$\m_\g=\r_\g e^{a|\g|}$ yields a strengthening of this condition that, however, has not been much used in practice.  The earlier work of Gruber and Kunz contained already an even better condition (but upper bounds on correlations were divergent at the
edge of convergence radius) .

\begin{theorem}{\bf \cite{GK}}\label{GKc}.
Let  $a>0$ and ${\bm \r}=\{\r_\g\}_{\g\in {\PP}}$ be collections of nonnegative numbers such that
$$
\sup_{x\in\VU}\sum_{\g\in  \PP\atop x\in \g}{\r_{\g}}\,\eee^{a\card\g}\;<\; e^{a}-1\;.\Eq(gks-r0)
$$
Then the functions \equ(diflog-g-sg) are analytic in the interior of the polydisc $\mathcal D_{\bm \r}$ and satisfy
$$
\card{\Theta}^{\L}_x(\bm\r)\;\le\; {1\over \eee^a}\, \biggl[ 1  + \sup_{x\in\VU}\sum\limits_{\g\in
\PP\atop x\in\g} \r_{\g}\,  \eee^{|\g|} \biggr]
\qquad
\Eq(gk.bo0)
$$
for all finite $\L\subset\VU$ and $x\in\L$.
\end{theorem}

Theorem \ref{FPn} yields a stronger result:

\begin{theorem}{\bf \cite{FP}}\label{FPc-r}.
Let  $\bm a=\{a_x\}_{x\in\VU}$ and ${\bm \r}=\{\r_\g\}_{\g\in {\PP}}$ be collections of nonnegative numbers such that
$$
\sum_{\g\in  \PP\atop x\in \g}{\r_{\g}}\,\eee^{\sum_{y\in\g}a_y}\;\le\; e^{a_x}-1\;.\Eq(gks-r)
$$
Then the functions \equ(diflog-g-sg) are analytic in the interior of the polydisc $\mathcal D_{\bm \r}$ and satisfy
$$
\card{\Theta}^{\L}_x(\bm\r)\;\le\; a_x
\Eq(gk.bo)
$$
for all finite $\L\subset\VU$ and $x\in\L$.
\end{theorem}

This result is obtained from \equ(fp)--\equ(fppn) through some rather rough approximations.  The argument is given in~\cite{FP} for the choice~\equ(choice), for completeness we show below the proof for general factorized weights.  The bound \equ(gk.bo) is the coarsest of a sequence of sharper bounds, as stated in Proposition \ref{pro:gks} below.

\\{\bf Proof.}
\\We shall prove that condition \equ(gks-r) implies \equ(fp) for
$$\mu_\g\;=\;\eee^{\sum_{x\in\g} a_x}\;.
\Eq(fact-2)
$$
With this choice the function \equ(fppn) becomes
$$
\varphi^{\rm FP}_\g(\bm \mu)\;=\;1+ \sum_{n= 1}^{|\g|}{1\over n!}
\sum_{(\g_1,\dots,\g_n)\in \PP^n} \prod_{i=1}^n\r_{\g_i} \eee^{\sum_{y\in \g_i} a_y}\prod_{1\le i<j\le n}\!\!\ind{\g_i\cap\g_j=\emptyset}
\prod_{i=1}^n\ind{\g_i\cap\g\neq \emptyset}
\Eq(fp-series)
$$
We now observe that a necessary condition to satisfy the indicator functions is the existence of \emph{different} points $x_i\in\g_i\cap\g$, $i=1,\ldots,n$ that each $\g_i$ intersect $\g$ at a different point.  Of course, the whole intersections must be disjoint but we only use the existence of \emph{some} set of different points. This approximation is reasonably if the contours involved are small and very bad otherwise.  In favorable case, this over-weighting of large contours may be masked by the smallness of the corresponding activities.  Therefore,
$$
\varphi^{\rm FP}_\g(\bm \m)\;\le\; 1+ \sum_{n=1}^{|\g|} \frac{1}{n!} \sum_{(x_1,\ldots,x_n)\in\g^n\atop x_i\neq x_j}
\prod_{i=1}^n\Biggl[\sum_{\g\in  \PP\atop x_i\in \g}{\r_{\g}}\,\eee^{\sum_{y\in\g} a_y}\Biggr]\;.
$$
Applying hypothesis \equ(gks-r) we obtain
\begin{eqnarray*}
\varphi^{\rm FP}_\g(\bm \mu) &\le& 1+ \sum_{n=1}^{|\g|}\sum_{\{x_1,\ldots,x_n\}{\subset}\g}\;
\prod_{i=1}^n\Bigl[\eee^{a_{x_i}}-1\Bigr]\\
&=& \prod_{x\in\g} \Bigl[(\eee^{a_{x}}-1) +1\Bigr] \;=\; \eee^{\sum_{x\in\g}a_x},
\end{eqnarray*}
and so
$\r_\g\varphi^{\rm FP}_\g(\bm \mu)\le \r_\g \eee^{\sum_{x\in\g}a_x}=\m_\g$
showing that \equ(fp) holds.  Finally, bound \equ(bPia) implies
$$
\card{\Pi}^{\L}_x(\bm\r)\le \varphi^{\rm FP}_{\{x\}}\;\le\;\eee^{a_x}
$$
so using \equ(rela) we get the bound \equ(gk.bo).
$\qed$


%

%
\subsection{Comments and overview of methods and results}

The preceding criteria suggested us a number of comments and questions that we answer in the sequel.

\begin{itemize}
\item[(C1)]  While Gruber-Kunz criterium improves the ``naive'' Dobrushin criterion obtained by the application of \equ(do) to subset gases, the actual condition \equ(gks-r) is the natural analogue of \equ(r.dob.1) under the replacement $\sum_{\gt\in\nbhd\g}$ by $\sum_{\gt\ni x}$.  This suggests that a correspondingly adaptated Dobrushin argument could lead to the same result.  This is true, as discussed in Section \ref{s:Dob}.  In fact such an argument proves the strongest criterion of Theorem~\ref{FPc-r}.

\item[(C2)]  The difference between convergence criteria of Theorems~\ref{FPc-r} and~\ref{GKc} looks small indeed.  But the non-sharp inequality is clearly out of reach of the Gruber-Kunz treatment of Kirkwood-Salzburg equations, based on establishing \emph{strict} contractions.  Is this an inherent limitation of studies based on such equations?  The answer is no, as we discuss in Section~\ref{s:GK}:  The same equations processed in a different way do yield a proof of Theorem~\ref{FPc-r}.

\item[(C3)]  Having proved that Dobrushin's and modified Gruber-Kunz approaches yield the same results for subset gases, it is natural to wonder whether this equivalence extends to general abstract polymers.  The answer is yes, and we show this in Section~\ref{s:KS}.

\end{itemize}
There is an aspect, however, in which the modified GK-approach excels Dobrushin.  The former leads to a whole sequence
of successively better bounds for the ratios of partition functions, improving \equ(r.dob.2) or \equ(gk.bo).
See Propositions \ref{pro:gks} and \ref{pro:fps}.  A similar hierarchy of bounds appears in our general approach based on the Penrose identity~\cite{FP}.

The discussion that follows will clearly show the key point of contact between
Dobrushin induction argument and Kirkwood-Salzburg  equations.  For the case of subset polymers, both rely on the \emph{site-addition identity}
$$
 \Xi_{Y\cup\{x\}}(\bm z)  \;=\;  \Xi_{Y}(\bm z)~
+~ \sum_{S\subset Y\atop |S|\ge 0} z_{\{x\}\cup S} \,\Xi_{Y\backslash S}(\bm z)
\Eq(GKin3)
$$
valid for any $Y\in \PP$ and any $x\in\VU\backslash Y$.  This identity follows immediately from definition \equ(Xis).
In the general abstract setting there is an analogous \emph{polymer-addition} identity consequence of \equ(2):
$$
\Xi_{Z\cup \g_0}(\r)= \Xi_{Z}(\bm z)+z_{\g_0}\,
\Xi_{Z\backslash\nbhdst{\g_0}}(\bm z) \Eq(fungk)
$$
valid for any finite family $Z\subset \PP$  and any polymer $\g_0\in \PP\backslash Z$.  Here $\nbhdst{\g_0}$ denotes the
\emph{punctured neighborhood} of $\g_0$:
$$
\nbhdst{\g_0}\;=\; \nbhd{\g_0}\backslash \{\g_0\}\;.
$$

Identity \equ(fungk), called ``the fundamental identity" by Scott and Sokal  (\cite{SS}, section 3.1, formula (3.3))
is, as explained there, the key point of inductive proofs \emph{\`a la} Dobrushin.  The Kirkwood-Salzburg equations set up by Gruber and Kunz, on the other hand, follow from a rewriting of this fundamental identity.  It is no surprise that both methods yield equivalent results.

\section{Induction method for the subset gas}
\label{s:Dob}
\numsec=3\numfor=1
Let us start by proving Theorem~\ref{FPc-r} \emph{\`a la} Dobrushin.  As in the original Dobrushin argument, the proof is amazingly short.
From identity \equ(minus) [consequence of the alternating-sign property] and the definition of $\Theta^\L_x$ [first line of \equ(diflog-g-sg)] we see that the theorem is equivalent to the following proposition.
\begin{proposition}
Let  $\bm a=\{a_x\}_{x\in\VU}$ and ${\bm \r}=\{\r_\g\}_{\g\in {\PP}}$ be collections of nonnegative numbers such that
$$
\sum_{\g\in  \PP\atop x\in \g}{\r_{\g}}\eee^{\sum_{y\in\g}a_y}\;\le\; \eee^{a_x}-1\Eq(gks-dob)
$$
then
$$
{\Xi_{\L\backslash \{x\}}(-\bm\r)\over \Xi_{\L}(-\bm\r)}\;\le\; \eee^{a_x}
\Eq(zzz)
$$
for any finite $\L\subset\VU$ and any $x\in\L$.
\end{proposition}

\\{\bf Proof.}
\\We proceed by induction on $\card\L$.
If $\card\L=1$, $\L=\{x\}$, hypothesis \equ(gks-dob) implies that
$$
\r_{\{x\}}\,\eee^{a_x} \;\le\; \sum_{\g\in  \PP\atop x\in \g}{\r_{\g}}\,\eee^{\sum_{y\in\g}a_y}\;\le\; \eee^{a_x}-1\;.
$$
Hence,
$$
{\Xi_{\{x\}\backslash \{x\}}(-\bm\r)\over \Xi_{\{x\}}(-\bm\r)} \;=\;
\frac{1}{\Xi_{\{x\}}(-\bm\r)} \;=\; \frac{1}{1-\r_x}
\;\le\; \frac{1}{1-(1-\eee^{-a_x})} \;=\; \eee^{a_x}\;.
$$
Assume \equ(zzz) is true for $\card\L\le n$.  Telescoping we conclude that
$$
{\Xi_{\L\backslash S}(-\bm\r)\over \Xi_{\L}(-\bm\r)}\;\le\; \eee^{\sum_{y\in S}a_y}
\Eq(sss)
$$
for any $S\subset\L$. Take $x\not\in\L$. The site-addition identity \equ(GKin3) implies
$$
{\Xi_{\L\cup\{x\}}(-\bm\r)\over \Xi_{\L}(-\bm\r)} \;=\; 1 -\sum_{S\subset
\L}\r_{\{x\}\cup S} \,{\Xi_{\L\backslash S}(-\bm\r)\over \Xi_{\L}(-\bm\r)}
$$
which by \equ(sss) yields
\begin{eqnarray*}
{\Xi_{\L\cup\{x\}}(-\bm\r)\over \Xi_{\L}(-\bm\r)} &\ge& 1 -\sum_{S\subset
\L}\r_{\{x\}\cup S} \,\,\eee^{\sum_{y\in S}a_y}\\
&=& 1 -\eee^{-a_x} \sum_{\g\in \PP : x\in\g}\r_{\g} \,\eee^{\sum_{y\in \g}a_y}
\end{eqnarray*}
Finally, using hypothesis \equ(gks-dob),
$$
{\Xi_{\L\cup\{x\}}(-\bm\r)\over \Xi_{\L}(-\bm\r)} \;\ge\;
1-\eee^{-a_x}(\eee^{a_x}-1) \;=\; \eee^{-a_x}\;.
$$
Hence \equ(zzz) holds for regions $\L$ with $n+1$ sites.
$\Box$
\bigskip

We basically adapted the version of Miracle-Sol\'e~\cite{Mi}, who used identity \equ(minus) to simplify the original proof of Dobrushin which did not resort to such identity.

\section{Gruber-Kunz formalism for the subset gas}
\label{s:GK}
\numsec=4\numfor=1

In this section we revise the steps followed by Gruber and Kunz in \cite{GK} to arrive
to their Theorem~\ref{GKc} and present the modifications needed to arrive to Theorem~\ref{FPc-r}.
In fact, the argument of Gruber and Kunz was presented in terms of a ``gas of partitions'' which basically corresponds to a gas of subsets with single-site fugacities equal to one.  We have transcribed it to the framework of general subset gases.

\subsection{The proof by Gruber and Kunz}
Gruber and Kunz obtained their analyticity results by setting up linear equations
for the reduced correlations $\bar\phi_\L(\bm z, X)$
[defined by the subset version of \equ(sveg2-r)] , involving a
$\L$-\emph{independent} operator $K$.
To find these equations Gruber and Kunz applied the so called ``algebraic method"
following  closely Section 4.4. of~\cite{Ru}. However, in the context of the subset gas,
due to the fact that $\bar\phi_\L(\bm z,X)$ are just  ratios of partitions functions,  their equations (\cite{GK} eq.~(28) pag.~146)
can be  derived much more easily as follows.
We start with the site-addition identity \equ(GKin3) written rather as a site-deletion identity, in the form
$$
\Xi_{\L\backslash X}(\bm z)  \;=\; \Xi_{\L\backslash(
X\backslash\{x_1\})}(\bm z)~ -~ \sum_{S\subset \L\backslash X\atop |S|\ge 0}
z_{\{x_1\}\cup S} \,\Xi_{\L\backslash(X\cup S)}(\bm z)\Eq(GKin4)
$$
valid for any finite $\L\subset\VU$ and any $x_1\in X\subset\L$.
Thus, upon dividing both sides by $\Xi_{\L}(\bm z)$,
$$
\bar\phi_\L(\bm z,X)\;=\; \bar\phi_\L(\bm z, X\backslash \{x_1\})~ -~
\sum_{S\subset \L\backslash X\atop |S|\ge 0} z_{\{x_1\}\cup S}\,
\bar\phi_\L(\bm z,X\cup S)\Eq(GKin)
$$
These are what Gruber and Kunz call the
\emph{Kirkwood-Salzburg equations for the gas of subsets}. In these
equations $x_1$ is some point of $X$ chosen once and for all, for instance as the smallest site of $X$ in some fixed enumeration of $\VU$.

In order to write this in terms of a $\L$-independent operator, it is necessary to include the restriction $X\subset\Lambda$ as a factor, so to extend the functions $\bar\phi_\L(\bm z,X)$, defined only when$X\subset \L$ to all $X\in \VU$.
Let us then define
$$
\chi_\L(X)\;=\;\ind{ X\subset \L}\Eq(chik)
$$
and denote
$$
\widetilde\phi_\L(\bm z,X)\;=\;\chi_\L(X)\,\bar\phi_\L(\bm z,X)\;.\Eq(tilfi)
$$
From  \equ(GKin) we obtain
$$
\widetilde\phi_\L(\bm z,X)  \;=\; \left\{\begin{array}{c}
\displaystyle\chi_\L(X)\widetilde\phi_\L(\bm z,X\backslash \{x_1\})\\
\displaystyle \chi_\L(X)
\end{array}\right\}
 - \chi_\L(X) \sum_{S\in \PP^*\atop S\cap X=\0}
\r_{\{x_1\}\cup S}\, \widetilde\phi_\L(\bm z,X\cup S) \Eq(tilphi)
$$
where $\PP^*=\PP\cup\0$ and
the upper line holds when $\card X \ge 2$ while the lower one when $\card X=1$.
The latter includes the condition  $\bar\phi_\L(\bm z,\emptyset)=1$ which is better written as an inhomogenity of the linear system by introducing
$$
\a(X)\;=\;\ind{X|=1}\;.\Eq(alpha)
$$
In this way we conclude that the function
$$
\widetilde\phi_\L(\bm z)(\,\cdot\,) \equiv \widetilde\phi_\L(\bm z,\,\cdot) : \PP^* \longrightarrow \mathbb C
$$
satisfies the linear equation
$$
\widetilde\phi_\L(\bm z)= \chi_\L \a + \chi_\L K_{\bm z} \,\widetilde \phi_\L (\bm\r) \Eq(KSE)
$$
where $K_{\bm z}$ is the linear operator on the space of complex-valued functions on $\PP$ defined by
$$
(K_{\bm z} f)(X)= \ind{|X|\ge 2}f(X\backslash\{x_1\}) - \sum\limits_{S\in P^* \atop
S\cap X=\0} z_{\{x_1\}\cup S}\, f(X\cup S)\;.\Eq(Kro)
$$

At this point Gruber and Kunz resort to a contraction argument in Banach spaces.  For this ---having in mind weights of the form \equ(fact)--- they associate to each family $\bm \xi=\{\xi_x\}_{x\in\VU}$, with each $\xi_x>0$, the Banach space $ \mathcal{B}_{\bm \x}$ of complex functions defined on non empty
finite subsets of $\VU$ (i.e. on { $\PP$}) with the norm
$$
\|f\|_{\bm \x}=\sup_{X\in \PP} {|f(X)|\over \bm\x^{X}}\;.
\Eq(bxi)
$$
{ where, for any $X\subset \VU$,  we  abbreviate $\bm\x^{X}=\prod_{x\in X}\x_x$.}

We have,
$$
\begin{minipage}{8cm}
\begin{eqnarray*}
\bigl|(K_{\bm z} f)(X)\bigr| & \le &
 \bm\x^{X\backslash\{x_1\}}\, \|f\|_{\bm \x}  + \sum\limits_{S\in P^* \atop S\cap X=\0}
\bigl|z_{\{x_1\}\cup S}\bigr| \, \bm \x^{X\cup S}\, \|f\|_{\bm \x}\\
& = & \bm\x^{X} \,\|f\|_{\bm \x} \,\, {1\over \x_{x_1}}\, \biggl[ 1  + \sum\limits_{\g\in \PP\atop x\in\g} \bigl|z_{\g}\bigr|\,  \bm \x^{\g}\;.
\biggr]
\end{eqnarray*}
\end{minipage}
\Eq(kzkz)
$$
Therefore $K_{\bm z}$ is a bounded operator in $ \mathcal{B}_\x$ with norm
bounded by
$$
\|K_{\bm z} \|_{\bm\x}\;\le\; \sup_{x\in\VU}{1\over \x_x}\, \biggl[ 1  + \sup_{x\in
\VU}\sum\limits_{\g\in \PP\atop x\in\g} |z_{\g}|\,  \x^{|\g|}
\biggr] \;.\Eq(normK)
$$


If $\bm z$ is such that
$$ \|K_{\bm z}\|_{\bm\x}<1\;,  \Eq(norm)
$$
the equation \equ(KSE) has a unique solution in the Banach space $ \mathcal{B}_{\bm\x}$ given by
$$
\widetilde \phi_\L(\bm z)= \bigl[1- \chi_\L K_{\bm z}\bigr]^{-1} ~\chi_\L \a \Eq(line)
$$
By construction, this solution is analytic in $\bm z$ and furthermore
$$\|\widetilde \phi_\L(\bm z) \|_{\bm\x}\le (1-\|K_{\bm z}\|_{\bm\xi})^{-1}\;.
\Eq(bbb)
$$
As the condition \equ(norm) is independent of $\L$ the equation \equ(line) makes sense in the limit $\L\to\VU$ and yields the convergence  $\widetilde \phi_\L(\bm z)\to \phi(\bm z)$ where the latter is the unique solution of \equ(line) without the factors $\chi_\L$.  Choosing $\x_x=\eee^a$ we see that the condition
$$
{1\over \eee^a}\, \biggl[ 1  + \sup_{x\in\VU}\sum\limits_{\g\in
\PP\atop x\in\g} \r_{\g}\,  \eee^{|\g|} \biggr]<1 \Eq(GKcondi)
$$
implies the validity of all these properties for $\bm z\in\mathcal D_\r$ for all $\g\in\PP$, plus analyticity in the interior of $\mathcal D_\r$.

\subsection{Extended Gruber-Kunz criterion}
To extend the convergence region so to include equality in \equ(GKcondi) { and to improve bound \equ(gk.bo0)} we have to abandon the precedent contraction strategy
and find an alternative way to make sense of \equ(line).  In fact, this expression corresponds to the multivariate formal power series in $\bm z$
$$
\chi_\L \sum_{n\ge 0}  \bigl[ \bigl(K_{\bm z}\bigr)^n (\chi_\L\a)\bigr](X)\;.\Eq(series)
$$
Thus, by the reasons invoked above \equ(GKcondi), it is enough to find a
polydisc $\mathcal D_\r$ ---independent of $\Lambda$ and $X$--- where all these series converge uniform and absolutely.
A glimpse at the definition \equ(Kro) of the operators $K_{\bm z}$
shows that if $\card{z_\g}\le \r_\g$ for all $\g\in \PP$, each series \equ(series) is term-by-term dominated by the series with positive terms
$$
\Phi_{\bm \r}(X)\;=\; \sum_{n\ge 0}    \bigl[ \bigl(K_{-\bm \r}\bigr)^n \,\a\bigr](X)\;.\Eq(seriep)
$$
In particular,  the reduced correlations satisfy
$$
\bigl|\bar \phi_\L(\bm z,X)\bigr|\;\le\; \phi_\L(\bm {-\r},X)
\;=\; {\Xi_{\L\backslash X}(-\bm\r)\over \Xi_{\L}(-\bm\r)}
\;\le\; \Phi_{\bm \r}(X)
\Eq(fibo)
$$
for all finite $\L$ and all $X\subset\L$, $x\in\L$ and $\card{z_\g}\le \r_\g$.
To prove Theorem \ref{FPc-r} we only need to find a $\L$- and $X$-independent family $\{\r_\g>0\}_{\g\in\PP}$ for which this series is finite.   This is done in the following proposition which yields some further bonds.

\begin{proposition}\label{pro:gks}
Let  $\bm \xi=\{\xi_x\}_{x\in\VU}$ and ${\bm \r}=\{\r_\g\}_{\g\in {\PP}}$ be collections of nonnegative numbers such that
$$
\sum_{\g\in \PP\atop x\in \g} \r_\g\, \bm \x^{\g}\;\le\; \xi_x-1  \Eq(dobpp)
$$
for all $x\in\PP$.  Then the reduced correlations are analytic in the interior of the poly-disc  $\mathcal D_{\bm\r}=\{|z_\g|\le \r_\g:\g\subset \L\}$
and satisfy the uniform bound
$$
\left|{\Xi_{\L\backslash X}(-\bm z)\over \Xi_{\L}(-\bm z)}\right| \;\le\; \bm\x^X
\Eq(unif)
$$
for all finite $\L$, all $X\subset\L$ and all $\bf z\in\mathcal D_\r$. Furthermore, this bound can be systematically improved in the following way.
Consider the operator $\mathbb T_{\bm\r}$ on functions $F$ on $\PP$ defined by
$$
\bigl(\mathbb T_{\bm\r} \,F\bigr)(X)\;\equiv\;(\a +  K_{\bm -\r} \,F)(X)\;.
\Eq(ttt)
$$
Then, for all $m\le n$
$$
\left|{\Xi_{\L\backslash X}(-\bm z)\over \Xi_{\L}(-\bm z)}\right|\;\le\;
{\Xi_{\L\backslash X}(-\bm\r)\over \Xi_{\L}(-\bm\r)} \;\le\;
\bigl(\mathbb T_{\bm\r}\bigr)^m \,\bm \x^X\;\le\;
\bigl(\mathbb T_{\bm\r}\bigr)^n \,\bm \x^X\;\le\; \bm\x^X
\Eq(seq.bo)
$$
for all finite $\L$, all $X\subset\L$ and all $\bf z\in\mathcal D_\r$.
\end{proposition}

\\{\bf Proof.}
\medskip

\noindent

We start by observing that the positivity of the coefficients involved in the definition of
$\mathbb T_{\bm\r}$ implies that
$$
F(X)\le G(X)\; \forall X\in \PP \quad\Longleftrightarrow\quad
\bigl(\mathbb T_{\bm \r}\, F\bigr)(X)\le \bigl(\mathbb T_{\bm\r}\, G\bigr)(X)\; \forall X\in \PP\;.
\Eq(monoton)
$$
Furthermore, for every non-negative function $\x(X)$ and every positive integer $k$,
$$
\sum_{n= 0}^k  \bigl[ \bigl(K_{\bm {-\r}}\bigr)^n \,\a\bigr](X) \;\le\;
\bigl(\mathbb T^{k+1}_{\bm\r} \x\bigr)(X)\;.
\Eq(iter)
$$
\smallskip

\noindent {\bf Claim 1:} \emph{The series $\Phi_\r(X)$ converge if and only if there exists a function $\x(X):\PP^* \longrightarrow [0,\infty)$ such that}
$$
\bigl(\mathbb T_{\bm\r} \,\x\bigr)(X)\;\le\; \x(X) ~~~~~~~~~~~~~~~~\forall X\in \PP\;.\Eq(KSEp)
$$
\smallskip

\noindent
 Indeed, sufficiency follows from the fact that, by \equ(iter)
 and the monotonicity property \equ(monoton),
$$
\Phi_\r(X)\;\le\; \lim_{k\to\infty}\bigl(\mathbb T^k_{\bm\r} \,\x\bigr)(X)\;\le\;\x(X)\;.
\Eq(denote)
$$
On the other hand, if the series $\Phi_\r(X)$ converge, then \equ(KSEp) is satisfied ---as equality--- with $\xi(X)=\Phi_\r(X)$.
\bigskip

\noindent {\bf Claim 2:}
\emph{If the family $\bm\x=\{\x_x\}_{x\in\VU}$ satisfies \equ(dobpp), then the functions $\xi(X)=\bm\x^X$ satisfy \equ(KSEp). }
 \smallskip

\noindent
We need only to check \equ(KSEp) for $\card X\ge 2$.  In this case
$$\begin{array}{rcl}
(\a +  K_{\bm -\bm\r} \;\x)(X)&=  &\x\bigl(X\backslash\{x_1\}\bigr) +
\sum\limits_{S\in P^* \atop
S\cap X=\0} \r_{\{x_1\}\cup S}\,\x(X\cup S)\\[20pt]
&= &\bm\x^{X\backslash\{x_1\}} \Bigl[1 + \sum\limits_{\g\in \PP^* \atop
x_1\in \g} \r_{\g}\,\bm\x^{\g}\Big]\;.
\end{array}\Eq(xK)
$$
As by \equ(dobpp) the last square bracket is less than $\xi_{x_1}$, the claim is proven.
\bigskip

Putting together the two claims we have proven the convergence of $\Phi_{\bm\r}$ whenever  condition \equ(dobpp) is satisfied.  As discussed above this yields analyticity in the interior of $\mathcal D_{\bm\r}$.  Successive applications of $\mathbb T_{\bm\r}$ to both sides of \equ(KSEp) yield, by the monotonicity property \equ(monoton) and  the leftmost inequality in \equ(denote),
the sequence of bounds
$$
\Phi_{\bm\r}(X) \;\le\;
\bigl(\mathbb T_{\bm\r}\bigr)^m \,\bm \x^X\;\le\;
\bigl(\mathbb T_{\bm\r}\bigr)^n \,\bm \x^X\;\le\; \bm\x^X\;.
$$
Due to the bound \equ(fibo) these inequalities prove  \equ(seq.bo). $\qed$

\section[Kirkwood-Salzburg formalism]{Kirkwood-Salzburg formalism for the abstract polymer gas}
\label{s:KS}
\numsec=5\numfor=1

To conclude, we show how the approach of the previous section can be adapted to prove Dobrushin criterion (Theorem \ref{Dob}) through Kirkwood-Salzburg equations.
As the treatment exactly parallels that for subset gases we shall only indicate the key expressions.

To derive the K-S equations we start from the polymer-addition identity~\equ(fungk) ---which is the same used by Dobrushin in his induction argument.  Upon dividing by $\L\supset Z\cup\{\g_0\}$ we obtain
$$
{\Xi_{Z\cup \g_0}(\bm z)\over \X_\L(\bm z)}= {\Xi_{Z}(\bm z)\over
\X_\L(\bm z)}+z_{\g_0} {\Xi_{Z\backslash\G^*(\g_0)}(\bm z)\over
\X_\L(\bm z)}\;. \Eq(absgk)
$$
Choosing $Z=\L\backslash X$ and writing $\G^*_{\L}(\g_0)=\G^*(\g_0)\cap \L$ we obtain that the reduced correlations \equ(sveg2-r) satisfy the equations
$$
 \bar \phi_\L(\bm z,X) \;=\; \bar \phi_\L(\bm z,X\backslash \g_0) - z_{\g_0} \,\bar \phi_\L\bigl(\bm z,X \cup\G^*_\L(\g_0)\bigr)\Eq(Ksgk)
$$
These are the Kirkwood-Salzburg equations for the abstract polymer gas.

As in the previous section we introduce $\chi_\L (X)=\ind{X\subset \L}$, $\a(X)=\ind{|X|=1}
$ and $\widetilde\phi_\L(\bm z, X)= \chi_\L (X)\bar\phi_\L(\bm z, X)$, so to write
\equ(Ksgk) in the form
 $$
\widetilde\phi_\L(\bm\r)= \chi_\L \a + \chi_\L\, K^{ \Lambda}_{\bm \r}\, \widetilde \phi_\L(\bm \r) \Eq(KSEs)
$$
with
$$
(K^{ \Lambda}_{\bm z} f)(X)= \ind{|X|\ge 2} f(X\backslash \g_0) - z_{\g_0} f(X
\cup\G_\L(\g_0))\Eq(opK)
$$
where $\g_0$ is the first polymer in $X$ in some previously chosen
enumeration.
{ Note also that now $K^\Lambda_{\bm z}$ depends also on $\L$ since
we recall that $\G^*_\L(\g_0)=\{\g\in \L: \g\not\sim\g_0\}$.  }
The goal is to make sense of
$$
\bigl[1- \chi_\L K^{ \Lambda}_{\bm z}\bigr]^{-1} ~\chi_\L \a \;=\;
\chi_\L \sum_{n\ge 0}  \bigl[ \bigl(K^{ \Lambda}_{\bm z}\bigr)^n (\chi_\L\a)\bigr](X)
\Eq(line-abs)
$$
{ simultaneously for all $\Lambda$. }

We can now transcribe \emph{exactly} the same steps as in the previous section with the notation
$$
\bm\x^X\;=\;\prod_{\g\in X}\x_\g\;.
\Eq(transc)
$$
In the original Gruber-Kunz approach the convergence of \equ(line-abs) is proven by showing that  $K_{\bf z}$ is a contraction on the space $\mathbb{B}_{\bm\x}$ of complex valued functions on polymers with norm $\|f\|_{\bm\x}$ defined as in \equ(bxi).  By a calculation completely analogous to \equ(kzkz) we obtain
$$
\|K^\L_{\bm z}\|_{\bm\x}\le \sup_{\g_0\in \PP}\biggl[{1\over
\x_{\g_0}}\Bigl[1+ |z_{\g_0}| \prod_{\g\in \G(\g_0)}\x_\g \Bigr]
\biggr] \Eq(normKs)
$$
With the substitution
$$\x_\g\;=\;\m_\g+1
\Eq(subst)
$$ this yields the convergence condition
$$
\sup_{\g\in\PP} \frac{\r_{\g}}{\m_{\g}}  \prod_{\gt\nsim\g} [1+\m_\gt]\;<\;1
$$
which is slightly weaker than Dobrushin condition \equ(do)--\equ(cridob).

To improve this condition we proceed as in the proof of Proposition~\ref{pro:gks} and focus rather on the convergence of the formal series defined by the right-hand-side of \equ(line-abs). A necessary and sufficient condition for this convergence is the existence of a function $\xi:\PP\longrightarrow \mathbb C$ such that
$$
(\a +  K_{-\bm \r} \,\x)(X)\;\le\; \x(X)\Eq(KSEpas)
$$
for every finite family $X\subset\PP$.
{ Here $K$ is the operator defined as in \equ(opK) but replacing $\G_\L(\g_0)$ by $\G(\g_0)$.}

$$
(K^{ \Lambda}_{\bm z} f)(X)= \ind{|X|\ge 2} f(X\backslash \g_0) - z_{\g_0} f(X
\cup\G_\L(\g_0))\Eq(opK2)
$$

Assuming the factorization hypothesis
\equ(transc) with $\xi_g>1$ we obtain,
as in \equ(xK),
$$
(\a +  K_{-\bm \r} \;\x)(X)\;\le  \;
\frac{\x(X)}{\x_{\g_0}} \Bigl[1+ \r_{\g_0}\,\x\bigl(\G(\g_0)\bigr)\Bigr]\;.
\Eq(wewe)
$$
This bound, combined with condition \equ(KSEpas) and the substitution \equ(subst) leads to the following proposition whose proof is a transcription of the proof of Proposition \ref{pro:gks}.

\begin{proposition}\label{pro:fps}
Let  $\bm \xi=\{\xi_\g\}_{\g\in\PP}$ and ${\bm \r}=\{\r_\g\}_{\g\in {\PP}}$ be collections of nonnegative numbers such that
$$
\r_{\g}\prod_{\gt\in \G(\g)}\x_\gt\;\le\; \x_{\g}-1  \Eq(dobpss)
$$
for all $\g\in\PP$.
Then the reduced correlations are analytic in the interior of the poly-disc  $\mathcal D_{\bm\r}=\{|z_\g|\le \r_\g:\g\subset \L\}$
and satisfy the uniform bound
$$
\left|{\Xi_{\L\backslash X}(-\bm z)\over \Xi_{\L}(-\bm z)}\right| \;\le\; \bm\x^X
\Eq(uniffp)
$$
for all finite $\L$, all $X\subset\L$ and all $\bf z\in\mathcal D_\r$. Furthermore, this bound can be systematically improved in the following way.
Consider the operator $\mathbb T_{\bm\r}$ on functions $F$ on $\PP$ defined by
$$
\bigl(\mathbb T_{\bm\r} \,F\bigr)(X)\;\equiv\;(\a +  K_{\bm -\r} \,F)(X)\;.
\Eq(tttfp)
$$
Then, for all $m\le n$
$$
\left|{\Xi_{\L\backslash X}(-\bm z)\over \Xi_{\L}(-\bm z)}\right|\;\le\;
{\Xi_{\L\backslash X}(-\bm\r)\over \Xi_{\L}(-\bm\r)} \;\le\;
\bigl(\mathbb T_{\bm\r}\bigr)^m \,\bm \x^X\;\le\;
\bigl(\mathbb T_{\bm\r}\bigr)^n \,\bm \x^X\;\le\; \bm\x^X
\Eq(seq.bofp)
$$
for all finite $\L$, all $X\subset\L$ and all $\bf z\in\mathcal D_\r$.
\end{proposition}
Note that  \equ(dobpss) becomes Dobrushin's criterion \equ(do) by substituting $\x_\g=1+\m_\g$.
\vv
\section{Conclusion}

The precedent arguments show that inductive DKP arguments and the use of KS equations are basically two alternative ways of exploiting the site-addition or polymer-addition identities \equ(GKin3) and \equ(fungk).  As the Kirkwood-Salzburg equations are exact relations between reduced correlations, they potentially include all the information needed to obtain successive improvements.  In fact, our analysis show where to aim: better bounds require better choices of functions $\xi(X)$ satisfying \equ(KSEpas). Such functions must not, therefore, be of the factorized form \equ(transc).  Expression \equ(wewe) implies the necessary condition
$$
 1+ \r_{\g_0}\,\x\bigl(\G(\g_0)\bigr)\;\le\;\x_{\g_0}
 \Eq(nec)
$$
for all $\g_0\in\PP$.  In fact, the condition found in~\cite{FP} is exactly of this form with
$\xi(X)=\Xi_X$.  We have been unable, however to prove the validity of \equ(KSEpas) for such a function $\xi$ for arbitrary $X$.

\section*{Acknowledgments}
This work  has been partially supported by
Conselho Nacional de Desenvolvimento Cient\'{i}fico e Tecnol\'ogico (CNPq),
 CAPES (Coordena\c{c}\~ao de Aperfei\c{c}oamento de Pessoal de
N\'{\i}vel Superior, Brasil) and FAPEMIG (Funda{c}\~ao de Amparo \`a Pesquisa do Estado de Minas Gerais). RB
thanks the hospitality of Laboratoire de Math\'{e}matiques
Rapha\"{e}l Salem in the Universit\'{e} de Rouen during his PhD work.
RF and AP  thank The Newton Institute for the generous support
during the Combinatorics and Statistical Mechanics programme.

\end{document}